# Carbon on the nanoscale: ultrastiffness and unambiguous definition of incompressibility

Almaz Khabibrakhmanov [a, b, c], Pavel Sorokin [a, b, c, *]

[a] *Technological Institute for Superhard and Novel Carbon Materials, 108840, 7a Tsentralnaya street, Troitsk, Moscow, Russian Federation*

[b] *National University of Science and Technology MISiS, 119049, 4 Leninskiy prospekt, Moscow, Russian Federation*

[c] *Moscow Institute of Physics and Technology (National Research University), 141701, 9 Institutskiy lane, Dolgoprudny, Moscow Region, Russian Federation*

## ABSTRACT

In the presented work, the features of mechanical stiffness of carbon nanoparticles (nanodiamonds and fullerenes) in a wide range of sizes are considered. The enhancement of nanodiamonds stiffness (comparing to bulk diamond) is studied and explained in the terms of average bond stiffness $\langle k \rangle_0$. It is shown that $\langle k \rangle_0$ can be useful in the description of various carbon nanostructures and gives reliable estimates of their incompressibility. Moreover, we found that $\langle k \rangle_0$ can be well estimated based only on relaxed atomic geometry.

## 1. Introduction

Diamond crystal is well-known for the variety of its extraordinary properties. In particular, diamond is famous for the unique mechanical stiffness having the highest bulk modulus and being referred to as the hardest crystalline material [1]. Search for superhard materials still draws close attention of materials scientists and attempts to find materials superior to diamond by stiffness, already counting successful studies of carbon nanotubes [2–4], ultrahard fullerites [5–7] and graphene [8,9], continue to this day.

Nanodiamonds (ND) represent small (several nm in size) diamond particles. From general considerations, it can be supposed that size effects in such particles should lead to modification of the original diamond properties. Structural, electronic and magnetic properties of nanodiamonds attract significant interest and have been studying for decades [10,11]. Special attention was also given to nanocrystalline diamond [12,13] due to its exceptional mechanical stiffness by which it can even exceed single crystal diamond. Nevertheless, the mechanical properties of single nanodiamonds – structural units of a nanocrystalline diamond – mainly remained unexplored until recently. However, the latest experimental evidence of nanodiamonds ultrastiffness [14,15] indicates that their mechanical properties also deserve detailed study.

---

[*] Corresponding author. Tel. +7 495 638-4415. E-mail: PBSorokin@tisnum.ru (Pavel Sorokin)

Conventionally, mechanical stiffness is characterized by bulk modulus $B_0$ as a measure of incompressibility. However, the calculation of elastic moduli for nanostructures can be puzzling due to the ambiguity of volume on the nanoscale. This problem was widely discussed for carbon nanotubes (CNTs) and fullerenes, and many different ways to determine their volume were proposed. Most often CNTs were described as hollow structures with a certain shell thickness $t$. The *ad hoc* convention is to set $t = 3.4$ Å [16,17] as a value of interlayer spacing in graphite. The alternative approach proposed by Yakobson [18] is based on continuum shell elasticity theory and gives the value of $t = 0.66$ Å. Later, many other approaches for finding shell thickness $t$ were examined [16,17,19]. As for fullerenes, Ruoff *et. al* were the first to estimate $B_0$ of $C_{60}$ molecule as 843 GPa [20] and 826 GPa [21]. In both cases, $C_{60}$ was considered as a homogenous elastic solid. With the same assumption, Peón-Escalante *et. al* [22] theoretically predicted $B_0 = 874$ GPa, but a spherical shell model of $C_{60}$ (with $t = 3.4$ Å) also considered in work [22] led to a significantly lower $B_0 = 640$ GPa.

Thus, there is an explicit problem with calculations of volume and, as a result, with the determination of elastic moduli for nanostructures, and alternative ways to characterize nanostructures mechanical properties should be considered. Here we suggest using average bond stiffness $\langle k \rangle_0$ as a mechanical characteristic of uniform hydrostatic compression. Actually, bond stiffness constants were often used before in studies of nanostructures mechanical properties, mainly to recalculate them into elastic moduli [21,23] with different assumptions about volume, as mentioned above. On the contrary, we propose to use $\langle k \rangle_0$ as a primary mechanical characteristic of isotropic compression on the nanoscale.

This paper is outlined as follows. In Sec.2, we define average bond stiffness and give details of *ab initio* calculations. The obtained results and discussions are presented in Sec.3. We introduce considered structures of nanodiamonds and fullerenes and briefly report their structural properties. Further, we discuss in details the features of nanodiamonds and fullerenes mechanical stiffness. We present the results for bulk modulus and average bond stiffness and compare these two approaches. We reveal the enhancement of both $B_0$ and $\langle k \rangle_0$ with decreasing size, but values of $B_0$ are shown to be dependent on the assumptions made about nanostructures volume. In contrast, we find that $\langle k \rangle_0$ is an unambiguous quantity that enables us to correctly compare various carbon nanostructures with each other as well as with crystalline materials by their mechanical rigidity. We also demonstrate that $\langle k \rangle_0$ could be reasonably estimated based only on relaxed atomic geometry.

## 2. Methods

### 2.1. Average bond stiffness

We define average bond stiffness in the following way:

$$\langle k \rangle = \frac{1}{N_b} \left( \frac{\partial^2 E}{\partial \langle l \rangle^2} \right). \tag{1}$$

Here $E$ is total energy, $N_b$ and $\langle l \rangle$ are the number of chemical bonds in a structure and their average length, respectively. This physical quantity has units of N m$^{-1}$ and does not require any ill-defined quantities, such as a thickness of atomic monolayer or nanocluster volume, to calculate its value. As mechanical properties of a covalently bonded solid are completely determined by the properties of its chemical bonds, values of bulk modulus $B$ and average bond stiffness $\langle k \rangle$ are not independent and related via the following equation (see the derivation in Supplementary Materials):

$$\langle k \rangle = \frac{V_0}{N_b \langle l \rangle_0^2} \left[ \frac{B}{1+\delta} \left( \frac{\partial \delta}{\partial \varepsilon} \right)^2 - P \left( \frac{\partial^2 \delta}{\partial \varepsilon^2} \right) \right], \tag{2}$$

where $\varepsilon = \frac{\langle l \rangle - \langle l \rangle_0}{\langle l \rangle_0}$, $\delta = \frac{V - V_0}{V_0}$ are a linear and volumetric strain, respectively. At zero pressure, it reduces to Eq. (3) that allows estimating bulk modulus based on average bond stiffness:

$$B_0^{est} = \langle k \rangle_0 \cdot \frac{N_b \langle l \rangle_0^2}{V_0 \left( \frac{\partial \delta}{\partial \varepsilon} \right)^2_{\varepsilon=0}}. \tag{3}$$

For two-dimensional materials one can introduce 2D equibiaxial strain ($\alpha$) and 2D analogs of pressure ($F$) and bulk modulus ($\gamma$, also referred to as layer modulus [24]): $\alpha = \frac{A - A_0}{A_0}$, $F = -\left( \frac{\partial E}{\partial A} \right)$, $\gamma = -A \left( \frac{\partial E}{\partial A} \right) = A \left( \frac{\partial^2 E}{\partial A^2} \right)$, where $A$ is a surface area. Average bond stiffness can be defined in the same way (1), as for bulk solids, and the relation between $\langle k \rangle$ and $\gamma$ is governed by the formula analogous to the Eq. (2):

$$\langle k \rangle = \frac{A_0}{N_b \langle l \rangle_0^2} \left[ \frac{\gamma}{1+\alpha} \left(\frac{\partial \alpha}{\partial \varepsilon}\right)^2 - F\left(\frac{\partial^2 \alpha}{\partial \varepsilon^2}\right) \right]. \tag{4}$$

*2.2. Computational details*

All calculations of the atomic structures in this work were done in the framework of density functional theory (DFT) [25,26] using the projector augmented wave (PAW) method [27,28], as implemented in VASP [29,30]. Electron exchange-correlation was treated in generalized gradient approximation as proposed by Perdew-Burke-Ernzerhof [31]. The plane waves energy cutoff was equal to 520 eV in all cases. Nanodiamonds and fullerenes were simulated as isolated nanoparticles surrounded by not less than 1 nm thick vacuum to avoid spurious interactions between periodic images of the structures. Calculations were done only at Γ-point of the corresponding Brillouin zone. Atomic coordinates relaxation was performed until interatomic forces converged to within 0.01 eV Å$^{-1}$. In the case of bulk diamond and graphene, the **k**-point sampling of Brillouin zone was done using 24×24×24 and 16×16×1 Monkhorst-Pack [32] meshes centered at Γ-point, respectively. For graphene, a vacuum space along Z-axis was set as large as 10 Å for the same purpose as in the case of nanoparticles.

### 3. Results and discussions

*3.1. Reference systems*

Bulk diamond and graphene were selected as references for mechanical properties calculations in this study due to a pure sp$^3$- and sp$^2$-hybridization realized in these materials, respectively. Therefore, the broad spectrum of the mechanical properties possible in covalently bonded carbon systems could be covered.

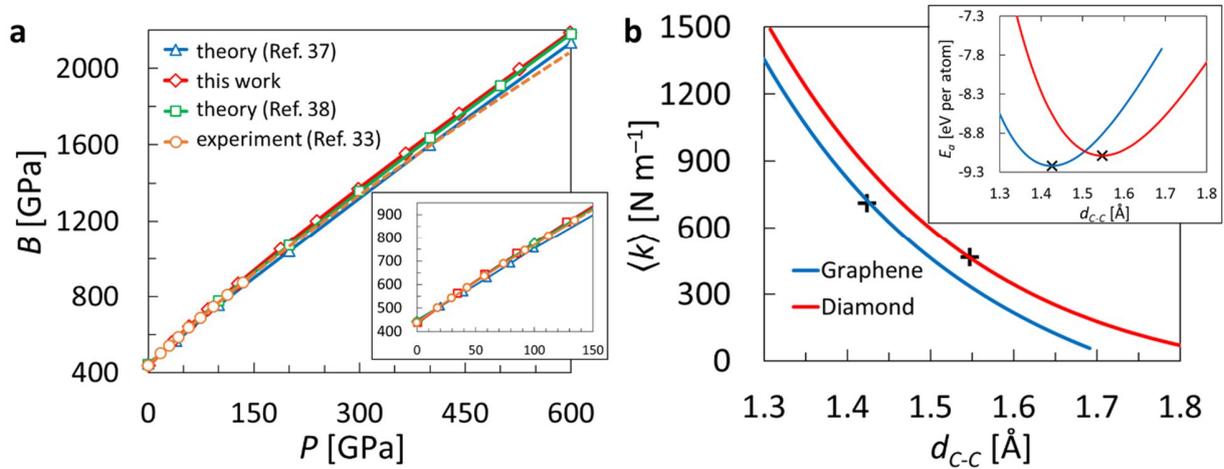

**Figure 1.** (a) The dependence of diamond bulk modulus on external pressure. The inset shows the low-pressure region in more details. Circles (up to 140 GPa) indicate experimental data [33], corrected for modifications to pressure calibration [34]. The dashed curve is an extrapolation of these data beyond 140 GPa based on the corresponding Vinet fit [34,35]. (b) The dependence of bond stiffness constant on bond length for graphene and bulk diamond. On the inset, energy per atom vs. bond length curves are drawn. Crosses denote the equilibrium state for each system.

We scrutinized the mechanical properties of diamond and graphene in a vast range of C–C bond lengths (1.3 – 1.8 Å). The obtained data were fitted to the corresponding 4$^{th}$ order equations of state (EOS) for bulk [36] and two-dimensional [24] materials, respectively. Extracted fitting parameters are given in Table 1. For diamond crystal, bulk modulus and pressure were found as derivatives of the energy with respect to the volume that allows to plot $B(P)$ dependence (Figure 1a). We compared our data with the previous theoretical [37,38] and experimental findings [33] and discovered a good agreement that verifies the obtained results. Then, the dependences of bond stiffness constant on bond length calculated using Eq. (2) and (4) were plotted (Figure 1b).

Although $\gamma_0$ and $\langle k \rangle_0$ are quantities very convenient for theoretical considerations, these cannot be directly measured that does not allow for direct comparison with experiment. Elastic moduli (in units GPa) are usually found instead. Therefore, we reviewed various theoretical as well as experimental works on elastic moduli of diamond and graphene (graphite) and recalculated elastic constants to $\gamma_0$ and $\langle k \rangle_0$ (see Table 1). For graphene, only $C_{11}$ and $C_{12}$ are required to estimate $\gamma_0$ [24] via $\gamma_0 = \frac{1}{2}(C_{11} + C_{12}) \cdot t$. Here $t$ is a thickness of graphene sheet which was set equal to the interlayer spacing of graphite given in the corresponding work. Then, $\langle k \rangle_0$ was found using Eq. (4) at zero strain: $\langle k \rangle_0 = \frac{4A_0\gamma_0}{N_b\langle l \rangle_0^2} = 2\sqrt{3}\gamma_0$. For diamond, $\langle k \rangle_0$ was directly estimated from bulk modulus via Eq. (3): $\langle k \rangle_0 = \frac{9V_0B_0}{N_b\langle l \rangle_0^2} = 3B_0a_0$. The data in Table 1 clearly indicate that our results are well-consistent with previous theoretical and experimental ones.

**Table 1.** Theoretical and experimental data on structural and mechanical characteristics of bulk diamond and graphene. Values obtained in the present work are given without superscript symbols. (Layer modulus $\gamma_0$ in N m$^{-1}$ and bulk modulus $B_0$ in GPa, their first derivatives $\gamma_0'$ and $B_0'$ dimensionless, their second derivatives $\gamma_0''$ and $B_0''$ in m N$^{-1}$ and GPa$^{-1}$, respectively).

|  | Graphene | | Diamond | |
| --- | --- | --- | --- | --- |
|  | theory | experiment | theory | experiment |
| Lattice constant $a_0$ [Å] | 2.468, 2.47[a], 2.446[b] | 2.459[c], 2.463[d], 2.462[e] | 3.572, 3.572[f], 3.574[g] | 3.567[h], 3.566[i], 3.575[j] |
| Layer modulus $\gamma_0$ of graphene and bulk modulus $B_0$ of diamond and their derivatives | $\gamma_0 = 207.6$ $\gamma_0' = 4.38$ $\gamma_0'' = -0.0398$ $\gamma_0 = 206.6$[a] | $\gamma_0 = 209.4$[d], 209.6[k], 207.7[l] | $B_0 = 434.6$ $B_0' = 3.84$ $B_0'' = -0.0116$ $B_0 = 433$[f] | $B_0 = 442$[m], 442[n], 444.5[i], 438[j] |
| Bond stiffness constant $\langle k \rangle_0$ [N m$^{-1}$] | 719.1, 715.7[a], 724.9[b] | 725.4[d], 726.1[k], 719.5[l] | 465.7, 464.0[f], 462.1[g] | 473[m], 475.5[i], 469.8[j] |

[a] Ref. [24]. [b] Ref. [39]. [c] Ref. [40]. [d] Ref. [41]. [e] Ref. [42]. [f] Ref. [43]. [g] Ref. [44]. [h] Ref. [45]. [i] Ref. [46]. [j] Ref. [33,34]. [k] Ref. [47]. [l] Ref. [48]. [m] Ref. [49]. [n] Ref. [50].

## 3.2. Considered structures

We investigated the mechanical properties of nanodiamonds and fullerenes in the wide range of sizes (54 – 1798 atoms). The set of studied nanodiamonds included 20 structures of cuboctahedral and cubic shapes. All structures were initially cleaved from the bulk diamond lattice, so their surfaces are composed of various low index diamond crystallographic planes. Considered geometrical shapes and the examples of nanodiamonds atomic structures are depicted in Figure 2. The cuboctahedral subset of studied nanodiamonds was represented by 5 structures ($C_{142}$, $C_{323}$, $C_{660}$, $C_{897}$, $C_{1201}$) faceted with 36% {111} surfaces and 64% {100} surfaces. The first cubic subset – cube {110} – contained 6 structures ($C_{54}$, $C_{259}$, $C_{509}$, $C_{712}$, $C_{1174}$, $C_{1798}$) encased with 67% {110} surfaces and 33% {100} surfaces, and the second one – cube {100} – involved 9 cubic nanodiamonds ($C_{75}$, $C_{239}$, $C_{387}$, $C_{568}$, $C_{577}$, $C_{812}$, $C_{920}$, $C_{1101}$, $C_{1465}$) enclosed entirely (100%) with {100} surfaces. The set of fullerenes comprised 7 molecules ($C_{60}$, $C_{180}$, $C_{320}$, $C_{540}$, $C_{720}$, $C_{960}$, $C_{1500}$), each of them belonging to the $I_h$ point symmetry group. In this study, fullerenes serve as an example of carbon nanoparticles with sp$^2$-hybridized bonds in contrast to sp$^3$-bonded nanodiamonds. To complete the picture, we also examined the structure and mechanical properties of nanodiamonds with the hydrogenated surface ($C_{268}H_{144}$, $C_{455}H_{196}$, $C_{660}H_{320}$, and $C_{712}H_{352}$).

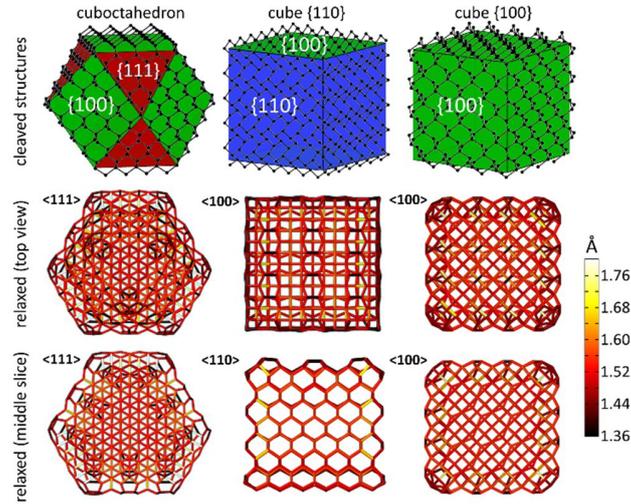

**Figure 2.** Zoo of studied nanodiamonds. Top row: Examples of cleaved atomic structures of nanodiamonds superposed onto the sketches of considered geometrical shapes (left to right): cuboctahedral and two families of cubic clusters with different faceting. For clarity, only surficial atoms are shown in the structures. Middle and bottom row: Relaxed atomic structures and color map of bond length distribution (top view and middle slice, respectively). View directions are indicated in the picture. This figure was created using OVITO [51] software.

All the structures were relaxed using DFT, as described in Sec.2.2. In the case of fullerenes, no dramatic changes in the atomic structure were observed after relaxation. The typical structure with 12 pentagons and $I_h$-symmetry was preserved. Little changes also occurred in hydrogenated nanodiamonds. Passivation of the surface by hydrogen atoms prevents it from the reconstruction and allows the bulk diamond structure to be preserved almost unchanged. Only small fluctuations in bond lengths and angles were observed in the relaxed structures. Therefore, they could be regarded as slightly distorted fragments of diamond crystal and should have mechanical properties similar to that of a crystalline diamond.

On the contrary, significant structural changes take place in bare nanodiamonds during relaxation. The presence of dangling bonds leads to the rearrangement of atoms on a surface, and peculiarities of this surface reconstruction strongly depend on surface morphology. Despite the presence of {111} surfaces in cuboctahedral nanodiamonds, their proportion is not very large, and {100} faces are predominant. This prevents cuboctahedral nanodiamonds from delamination of {111} planes, although bonds in this region elongate to 1.60 – 1.75 Å. Reconstruction of {100} surfaces in nanodiamonds occurs similar to the reconstruction of (100) surface in the bulk [52], primarily resulting in the formation of 2 × 1 dimer rows on the faces. In nanodiamonds from cuboctahedral and cube {100} subsets, characteristic fullerene-like caps [53] are present in the corners. These caps are composed of well-pronounced pentagons with the bond lengths (1.42 – 1.46 Å) and angles (~106º – 109º) that are almost identical to values observed in fullerenes (1.45 Å, 108º). The least structural changes were observed in nanodiamonds from cube {110} subset, where only moderate lattice distortions on {110} surfaces (in consistence with {110} bulk diamond surface reconstruction [54]) accompanied with the aforementioned formation of 2 × 1 dimers on {100} faces were found. The examples of relaxed nanodiamonds structures are shown in Figure 2. More details about the structural properties of nanodiamonds could be also found in the literature [55,56].

Generally speaking, the surface reconstruction that occurs in nanodiamonds results in the substantial transformations of the original $sp^3$-bonds of diamond. Some bonds in bare nanodiamonds turn into $sp^2$-like ones, whereas others remain closer to $sp^3$-bonds but in the strained state. However, there is a common trend towards a predominant presence of shortened (in comparison with the bulk) bonds over the lengthened ones, although the degree of predominance depends on the nanodiamonds' morphology. This trend could be clearly seen from color maps of bond length distribution in nanodiamonds given in Figure 2. We believe that the effect of bonds shortening in nanodiamonds has a key influence on their mechanical properties, to the discussion of which we turn.

*3.3. Mechanical stiffness*

To calculate nanostructures' mechanical stiffness, their relaxed atomic structures were subjected to small uniform isotropic deformations (were slightly expanded or compressed) with the following constrained geometry optimization to evaluate their total energy. The constraints are mandatory when performing simulations of the strained nanostructures because otherwise, they would simply go back to the relaxed state due to the absence of periodic boundary conditions on their surface. We found that it is enough to freeze only surficial carbon atoms that have fewer than 4 neighbors to keep structures in the stressed state, whereas all other atoms can be allowed to relax during optimization.

To be able to evaluate bulk modulus of the structure, one should determine its volume. However, there are various ways to determine volume on the atomic scale. In this study, we compared two main approaches to the calculation of nanostructures volume. In the first one, the structure is considered as a continuum medium bounded by an outer shell that has no thickness. The size of atoms is neglected, and nanostructure volume is calculated as volume enclosed by the convex hull of all atoms. For fullerenes, this coincides with the volume used by Ruoff *et. al* [20,21]. In the following text, we recall this definition as a solid medium volume ($V^{solid}$).

The second main approach implies finite size of atoms which can be taken into account in different ways. For nanodiamonds, we evaluated van der Waals volume ($V^{vdW}$). Atoms were imagined as spheres which can intersect and overlap, and $V^{vdW}$ was found as a total volume of these spheres. To compute $V^{vdW}$ we used the freeware ASV by Petitjean [57] and set the radius of carbon atoms to be equal $r_C = 1.7$ Å following the work of Bondi [58]. For fullerenes, we calculated an elastic shell volume ($V^{shell}$) as a volume of a hollow sphere with the thickness $t$ of its shell. This definition is very convenient for hollow nanostructures and allows for direct comparison by mechanical properties with graphene sheet (which is an infinite limit of fullerenes) having the same thickness $t$. As there is a wide scatter of wall thickness values for CNTs (0.66 – 3.4 Å), we find it possible to vary fullerenes shell thickness in the same range, as these carbon nanostructures are closely related. For briefness, we present results only for $t = 3.4$ Å in the main body of the article with others available in Supplementary Materials.

Calculations of bulk modulus were carried out for solid medium ($B_0^{solid}$), van der Waals ($B_0^{vdW}$) and elastic shell ($B_0^{shell}$) volume definitions using the unified methodology. The obtained energy vs. volume curves were approximated by the Birch-Murnaghan EOS [59,60] to retrieve $B_0$. Calculations with linear strains of $\varepsilon = \pm 0.015, \pm 0.03$ were performed in all cases as we found 5 points in the dataset (including relaxed structure) to be enough to gain accurate results. The acquired results are shown in Figure 3a. A significant stiffening of nanodiamonds in comparison with crystalline diamond is observed. This effect is present in both approaches to volume calculation and can be explained by the predominant presence of shortened (and therefore stiffened) bonds in the structures (Figure 2). Moreover, we found that the dependence of bulk modulus on nanostructures size has a hyperbolic character that also reflects the contribution of surficial atoms. This size dependence could be fitted by one hyperbola for all nanodiamonds families, that speaks of no significant influence of nanodiamonds shape on their mechanical stiffness.

Nevertheless, different volume definitions lead to discrepant bulk moduli, and in all cases $B_0^{vdW}$ values are higher than the values of $B_0^{solid}$. Indeed, volumetric deformations corresponding to $V^{vdW}$ are significantly lower than the deformations computed in terms of $V^{solid}$ that results in the steeper $E(\delta)$ curves (see Figure S1a in Supplementary Materials) and, as a consequence, leads to an increase of the calculated $B_0$. Moreover, for fullerenes regarded volume definitions lead to completely different trends. On the one hand, $B_0^{shell}$ tends to the constant value close to the effective bulk modulus of graphene (we defined this quantity as $B_0^{eff} = \frac{\gamma_0}{t} \approx 611$ GPa for $t = 3.4$ Å). On the other hand, monotonous fall via hyperbolic law is observed for $B_0^{solid}$ due to a significant volume of a void inside fullerene, which is not included in $V^{shell}$.

Thus, different approaches to the computation of volume result in different values of bulk modulus, and it is not clear in advance which values of $B_0$ one should use to characterize the rigidity of the structure. To overcome this problem, we propose an alternative approach of average bond stiffness $\langle k \rangle_0$. This physical quantity has an important advantage in comparison with bulk modulus, as it does not require to estimate structures volume. This relieves from uncertainties which we discussed above and makes the value of $\langle k \rangle_0$ unambiguous.

To calculate $\langle k \rangle_0$, we built bond lengths distribution at each step of deformation for all the structures to find the total number of bonds $N_b$ and their average length $\langle l \rangle$, and then Eq. (1) was used to retrieve $\langle k \rangle_0$. The obtained results are represented in Figure 3b. As it follows from the figure, values of $\langle k \rangle_0$ are scattered in the range between diamond and graphene bond stiffnesses. Let us point out that all studied nanodiamonds with bare surface overwhelm bulk diamond by the average bond stiffness that increases with decreasing their size, whereas $\langle k \rangle_0$ of hydrogenated nanodiamonds is lower than in crystal and does not manifest any size dependence. This is one more evidence showing that the nanodiamonds stiffness is completely determined by the properties of constituting chemical bonds. As for fullerenes, with their enlargement average bond stiffness tends to the constant value corresponding to graphene (Table 1). Thus, $\langle k \rangle_0$ correctly reproduces the main trends and limiting cases of the considered nanostructures stiffness.

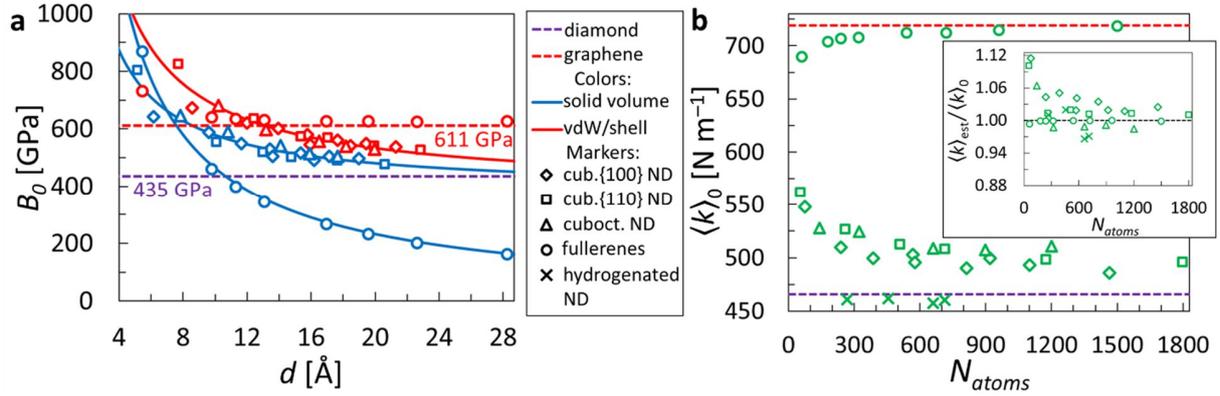

**Figure 3.** (a) Size dependence of considered nanostructures $B_0$ for different volume definitions which are shown in different colors (see the legend). Bulk moduli were retrieved from Birch-Murnaghan approximation of energy-volume curves. Various families of studied nanostructures are displayed with different markers. Hyperbolic fitting for each volume definition is also shown with a solid line of the corresponding color. The size $d$ of nanostructures was determined as a cubic root of the corresponding volume. (b) Average bond stiffness $\langle k \rangle_0$ vs. the number of atoms in nanostructures $N_{atoms}$. Purple and red dashed lines are given for reference and mark bond stiffness in bulk diamond and graphene, respectively. Ratio of $\langle k \rangle_0^{est}$ calculated based on bond lengths distribution to directly calculated stiffness $\langle k \rangle_0$ is drawn in the inset. The black dashed line denotes unity and is drawn to guide an eye. The legend in the figure applies both to frames (a) and (b).

The obtained results can be compared to experimental and theoretical data (Table 2). As one can see, experimentally measured values of nanodiamonds bulk modulus fall in the range predicted by our *ab initio* calculations. As there are no available experimental data on the compressibility of single fullerene molecules (at least, to our knowledge), we compare our results for $C_{60}$ with the previous theoretical works. Again, a good coherence in numbers is observed for both $B_0$ and $\langle k \rangle_0$ with the results of Ruoff *et. al* [21] and Peón-Escalante *et. al* [22].

**Table 2.** Values of $B_0$ and $\langle k \rangle_0$ for nanodiamonds and $C_{60}$ fullerene.

|  | Nanodiamonds | | Fullerene $C_{60}$ | |
| --- | --- | --- | --- | --- |
|  | present work | experiment | present work | previous theory |
| Bulk modulus $B_0$ [GPa] | 480 – 670 | 564[a], 607[b] | 868 | 826[c], 874[d] |
| Average bond stiffness $\langle k \rangle_0$ [N m$^{-1}$] | 485 – 562 | – | 689.9 | 672[c], 691[d] |

[a] Ref. [14]. [b] Ref. [15]. [c] Ref. [21]. [d] Ref. [22].

To gain a deeper insight into the nature of average bond stiffness as a mechanical characteristic of a nanostructure, we expressed it as a weighted sum over all bonds stiffness constants: $\langle k \rangle_0^{est} = \frac{\sum k_i}{N_b}$. The validity of this approach is justified by the localized character of covalent bonding in carbon materials (except for graphite and onions, where van der Waals bonding between the layers is present), that allows representing the total energy as a sum of energies per bond $E = \sum E_i$ and associating individual stiffness constant with each bond. The similar concept of atomic bulk modulus was successfully exploited by Kleovoulou *et. al* [61,62] in studies of silicon nanocrystals local rigidity and in our previous papers about polymerized fullerenes stiffness [6].

Relying on the number of neighbors, values of bond length and bond angles, we assigned the most appropriate by these structural parameters type of hybridization (sp$^2$ or sp$^3$) to each bond (technical details are presented in Supplementary Materials). Further, we used the dependence of bond stiffness on bond length for sp$^2$- and sp$^3$-hybridized carbon (Figure 1b) to calculate $k_i$ and then $\langle k \rangle_0^{est}$. Acquired results (see the inset in Figure 3b) demonstrate good agreement between $\langle k \rangle_0^{est}$ and $\langle k \rangle_0$ with most errors less than 5% which is decreasing with nanostructures enlargement. The fact that directly calculated $\langle k \rangle_0$ can be decomposed on a sum of individual bond stiffness constants directly indicates that the effect of nanodiamonds stiffening could be entirely explained and understood from the consideration of chemical bonds in the studied structures. This fact also confirms that the freezing of the surface atoms seems not to influence strongly the obtained values of $\langle k \rangle_0$, since $\langle k \rangle_0^{est}$ is not affected by any constraints and nevertheless, shows good agreement with $\langle k \rangle_0$. Moreover, the mentioned procedure gives an opportunity to estimate the value of the average bond stiffness having only relaxed atomic geometry of the structure. In principle, this relieves from the need for optimizations of strained structure geometry, with frozen surficial atoms or may be used to verify the results of such direct calculations. We suppose this to be an appealing advantage of our approach from the computational point of view.

Thus, $\langle k \rangle_0$ is a convenient and reliable quantity characterizing the mechanical properties of nanostructures locally. To find the relation between $\langle k \rangle_0$ and $B_0$ (which should describe stiffness of the structure as a whole), the Eq. (3) should be used. It enables to recalculate average bond stiffness into bulk modulus for a given relationship $\delta(\varepsilon)$ under hydrostatic compression conditions, which depends on the chosen definition of volume. For the volumes considered here, we have (see the derivation in Supplementary Materials):

$$B_0^{est} = \langle k \rangle_0 \cdot \frac{N_b \langle l \rangle_0^2}{9 V_0^{solid}}, \tag{5}$$

$$B_0^{est} = \langle k \rangle_0 \cdot \frac{N_b \langle l \rangle_0^2}{9 V_0^{vdW}} \cdot \left(1 + \frac{2r_C}{a_0}\right)^2, \tag{6}$$

$$B_0^{est} = \langle k \rangle_0 \cdot \frac{N_b \langle l \rangle_0^2}{4 V_0^{shell}}. \tag{7}$$

The results in Figure 4 show that the suggested method allows estimating nanostructures bulk modulus with an accuracy of 14% at worst and works much better in most cases. The error reduces with increasing nanostructures size, and thus reasonable estimates of $B_0$ can be obtained using Eq. (5) – (7). These equations are approximate, but in principle, the exact ones can be obtained using the same ideas. However, we suppose that this is not required for the purposes of a rough estimate, especially since bulk modulus is ambiguous on the nanoscale by its nature. We believe that the obtained results provide both qualitative as well as a quantitative understanding of the intrinsic relationship between bulk modulus and average bond stiffness. We also found that after a small modification the proposed method is capable of describing mechanical properties of defective structures (see Supplementary Materials).

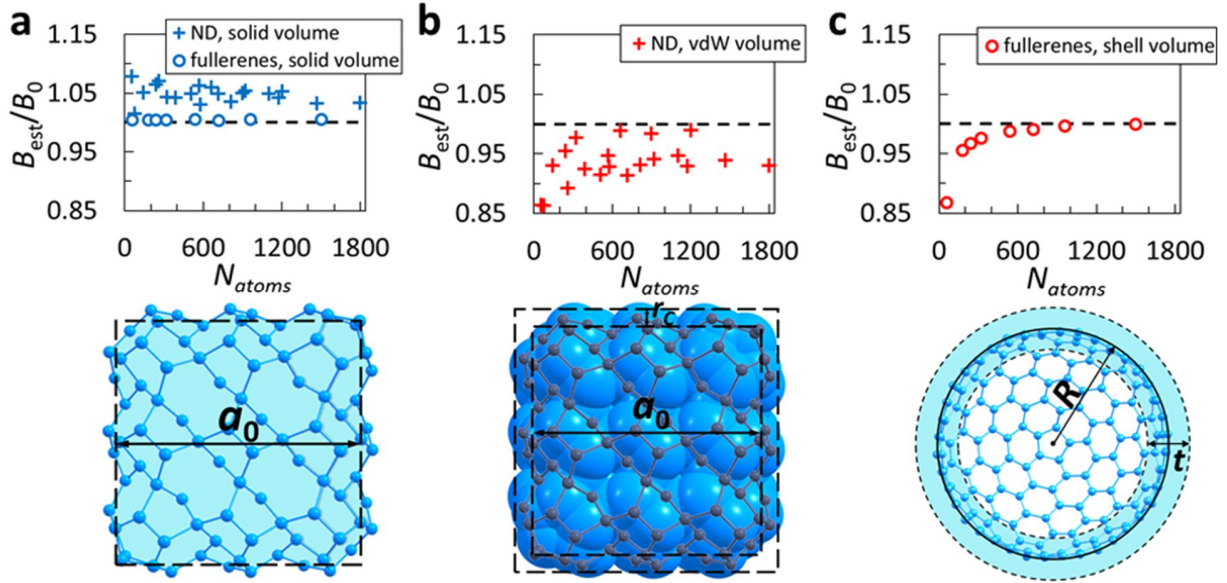

**Figure 4.** Ratio of $B_0^{est}$ to directly calculated bulk modulus $B_0$ for different volume definitions: (a) solid medium volume, (b) van der Waals volume, (c) elastic shell volume. Illustrations to the considered volume definitions are represented in the bottom.

However, the average bond stiffness approach also has its own limitations. By its definition postulating that strain energy is more or less uniformly distributed by individual chemical bonds in the structure, the usage of $\langle k \rangle_0$ for characterizing nanostructure stiffness is restricted to the case of covalently bonded solids. For instance, $\langle k \rangle_0$ cannot be applied for the description of layered nanostructures where van der Waals interactions play an important role (*e.g.*, carbon onions and onion-like structures). Indeed, the application of moderate external pressure to such structures should lead only to outer shell compression whereas inner core would remain almost undistorted. Therefore, the average bond stiffness approach would give much higher stiffness values that do not make any physical sense in such cases (see Supplementary Materials for additional discussion).

## 4. Conclusions

Based on our findings, we conclude that average bond stiffness should be considered as a primary characteristic of mechanical stiffness on the nanoscale. $\langle k \rangle_0$ enables one to unambiguously determine which structure is stiffer whether it is bulk crystal or low-dimensional object. This could be done basing only on relaxed atomic geometry that is the important advantage of this method. In principle, this approach could be applied to studies of any carbon nanostructures (*e.g.*, nanorods, nanoribbons), provided that they are covalently bound. Moreover, there are no restrictions for extending this method to nanostructures composed by other elements capable of forming covalent solids, such as Si, Ge, *etc*. Thus, there is a wide scope for further investigations, and we believe that our results could be useful for future works in this area.

**Acknowledgments**

This work was funded by an RFBR grant #18-29-19019. We also acknowledge the financial support of the Ministry of Education and Science of the Russian Federation in the framework of Increase Competitiveness Program of NUST "MISiS" (No. K2-2019-016) and Grant of President of Russian Federation for government support of young DSc. (MD-1046.2019.2). The calculations were performed at the supercomputer cluster provided by the Materials Modeling and Development Laboratory at NUST "MISiS". We are grateful to Prof. David Tománek for the fruitful discussions.

**Supplementary Materials**

Additional information such as derivations of equations used in the main text, auxiliary figures and extra discussions of some results, are included in the Supplementary Materials which are available online.